\documentclass[cits]{PoS}

\title{Critical properties of $2D$ $Z(N)$ vector models for $N>4$}

\ShortTitle{Critical properties of $2D$ $Z(N)$ vector models for $N>4$}

\author{Oleg Borisenko\\
        Bogolyubov Institute for Theoretical Physics, 
        National Academy of Sciences of Ukraine \\
        03680 Kiev, Ukraine \\
        E-mail: \email{oleg@bitp.kiev.ua}}

\author{\speaker{Gennaro Cortese}\\
        Instituto de F\'{\i}sica Te\'orica UAM/CSIC, 
        Cantoblanco, E-28049 Madrid, Spain,
        \\Departamento de F\'{\i}sica Te\'orica, 
        Universidad de Zaragoza, E-50009 Zaragoza, Spain
        \\Dipartimento di Fisica, Universit\`a della Calabria,
        and INFN - Gruppo Collegato di Cosenza \\I-87036 Rende, Italy\\
        E-mail: \email{cortese@cs.infn.it}}

\author{Roberto Fiore\\
        Dipartimento di Fisica, Universit\`a della Calabria,
        and INFN - Gruppo Collegato di Cosenza \\I-87036 Rende, Italy\\
        E-mail: \email{fiore@cs.infn.it}}

\author{Mario Gravina\\
        Department of Physics, University of Cyprus\\ 
        P.O. Box 20357, Nicosia, Cyprus\\
        E-mail: \email{gravina.mario@ucy.ac.cy}}

\author{Alessandro Papa\\
        Dipartimento di Fisica, Universit\`a della Calabria, 
        and INFN - Gruppo Collegato di Cosenza \\I-87036 Rende, Italy\\
        E-mail: \email{papa@cs.infn.it}}

\abstract{We investigate the critical properties of two-dimensional $Z(N)$
vector models for $N$ larger than 4. In particular, critical points of the
two phase transitions are located and some critical indices are determined. 
We study also the behavior of the helicity modulus and the dependence of the 
critical points on $N$.}

\FullConference{The XXIX International Symposium on Lattice Field Theory, 
Lattice2011\\July 11-16, 2011\\The Village at Squaw Valley, USA}

\begin{document}

\section{Introduction}
\label{intro}

The Berezinskii-Kosterlitz-Thouless (BKT) phase transition was originally 
discovered in the two-dimensional ($2D$) $XY$ model in the first half of 
the seventies~\cite{BKT}. 
Since then it was realized that this type of phase transition takes place in a 
number of other models including discrete $2D$ $Z(N)$ models for large enough 
$N$ and even $3D$ gauge models at finite temperature 
(see~\cite{3du1ft,3du1full} for a recent study of the deconfinement transition 
in $3D$ $U(1)$ lattice gauge theory). Here we are interested in the phase 
structure of $2D$ $Z(N)$ vector models. On a $2D$ lattice $\Lambda = L^2$ 
with linear extension $L$ and periodic boundary conditions, the partition 
function of the model can be written as
\begin{equation}
Z(\Lambda, \beta ) =\left[ \prod_{x\in \Lambda} 
\frac{1}{N} \sum_{s(x)=0}^{N-1} \right ]  \ \exp \left[ \sum_{x\in\Lambda} \ \sum_{n=1,2} \ 
\beta \ \cos\frac{2\pi }{N}( s(x)-s(x+e_n) ) \right]  \ .
\label{PFZNdef}
\end{equation}
The BKT transition is of infinite order and is characterized by the essential 
singularity, {\it i.e.} the exponential divergence of the correlation length. 
The low-temperature or BKT phase is a massless phase with a power-law decay of 
the two-point correlation function governed by a critical index $\eta$.  
The $Z(N)$ spin model in the Villain formulation has been studied analytically 
in Refs.~\cite{Villain}. It was shown that the model has at least two 
BKT-like phase transitions when $N\geq 5$.  The critical index $\eta$ has been 
estimated both from the renormalization group (RG) approach of the 
Kosterlitz-Thouless type and from the weak-coupling series for the
susceptibility. It turns out that $\eta(\beta^{(1)}_{\rm c})=1/4$ at the 
transition point from the strong coupling (high-temperature) phase to the 
massless phase, {\it i.e.} the behavior is similar to that of the $XY$ model. 
At the transition point $\beta^{(2)}_{\rm c}$ from the massless phase to 
the ordered low-temperature phase, one has $\eta(\beta^{(2)}_{\rm c})=4/N^2$. 
A rigorous proof that the BKT phase transition does take place, and so that the
massless phase exists, has been constructed in Ref.~\cite{rigbkt} for both 
Villain and standard formulations.
Monte Carlo simulations of the standard version with $N=6,8,12$ were performed 
in Ref.~\cite{cluster2d}. Results for the critical index $\eta$ agree well with
the analytical predictions obtained from the Villain formulation of the model.

In Refs.~\cite{z5_lat10,z5_phys.rev} we have started a detailed numerical 
investigation of the BKT transition in $2D$ $Z(N)$ models for $N=5$ which is 
the lowest number where this transition can occur. Our findings support the 
scenario of two BKT transitions with conventional critical indices. Here we 
continue our study with investigation of models for $N$=7 and 17. We want to 
locate the transition points and to compute some critical indices. Such 
results could serve as checking point of universality in our studies of the 
BKT transitions in $3D$ gauge models at finite temperature. 
Our second goal is to use the available data on the position of critical 
points to deduce phenomenological scaling of these points with $N$. 

\section{Numerical results}
\label{numerical}

We simulated the model defined by Eq.~(\ref{PFZNdef}) using the same cluster 
Monte Carlo algorithm adopted in the Refs.~\cite{z5_lat10,z5_phys.rev} for the 
case $N=5$.
We used several different observables to probe the two expected phase 
transitions. In order to detect the first transition ({\it i.e.} the one
from the disordered to the massless phase) we used the absolute value 
$|M_{L}|$ of the complex magnetization,
\begin{equation}
M_{L}=\frac{1}{L^{2}}\sum_{i}\exp\left(i\frac{2\pi}{N}s_{i}\right)\equiv |M_{L}|e^{i\psi},
\label{magn_complex}
\end{equation}
and the {\em helicity modulus}~\cite{hel_def,torsion} 
\begin{equation}
\Upsilon=\left<e\right> - L^{2}\beta\left<s^{2}\right>,
\label{Helicity_formula}
\end{equation}
where $e\equiv\frac{1}{L^{2}} \sum_{<ij>_{x}} \cos\left(\theta_{i}-\theta_{j}
\right)$ and $s\equiv\frac{1}{L^{2}} \sum_{<ij>_{x}} \sin(\theta_{i}
-\theta_{j})$.
For the second transition ({\it i.e.} the one from the massless to the ordered 
phase) we adopted the real part of the "rotated" magnetization, 
\[
M_{R}=|M_{L}|\cos(N\psi)\;,
\]
and the order parameter
\[
m_{\psi}=\cos(N\psi)
\]
introduced in Ref.~\cite{BMK09}, where $\psi$ is the phase of the complex 
magnetization defined in Eq.~(\ref{magn_complex}). 
In this work, both for $N=7$ and $N=17$, we 
collected typically 100k measurements for each value of the coupling $\beta$, 
with 10 updating sweeps between each configuration. To ensure  thermalization 
we discarded for each run the first 10k configurations. The jackknife method 
over bins at different blocking levels was used for the data analysis.

\begin{figure}
\begin{center}
\includegraphics[scale=0.25]{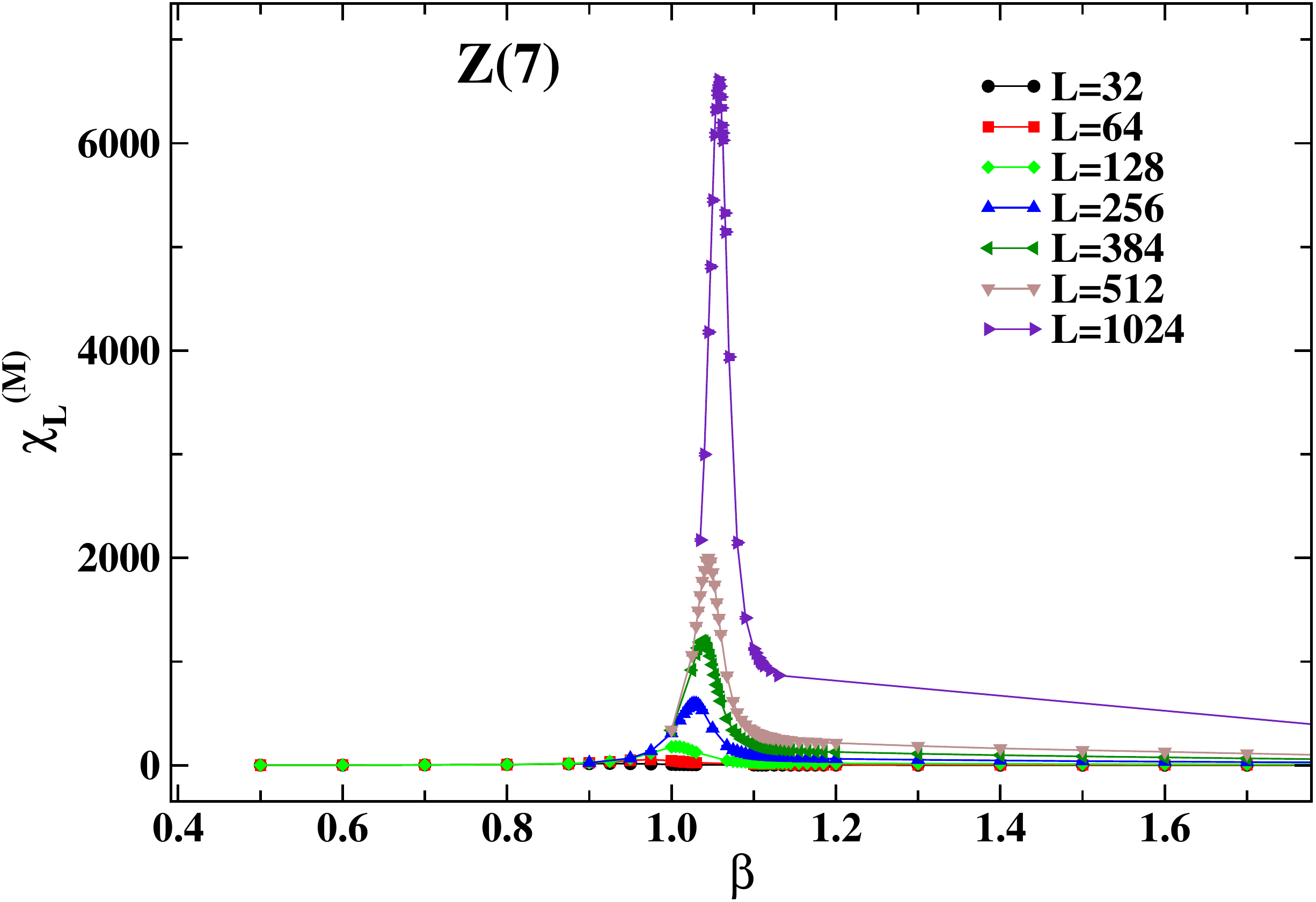}\hspace{0.5cm}
\includegraphics[scale=0.25]{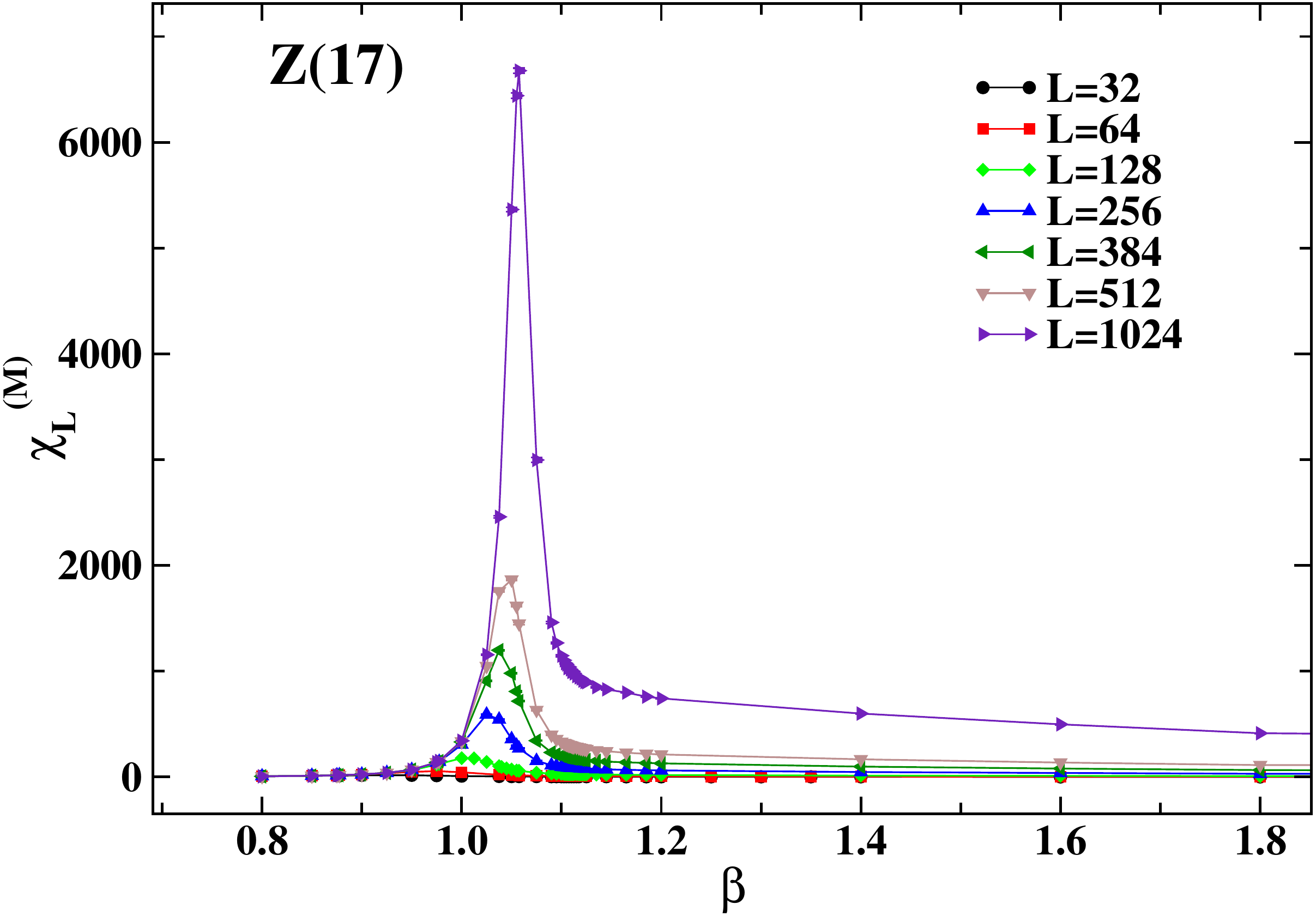}
\caption{Susceptibility $\chi_{L}^{(M)}$ versus $\beta$ 
in $Z(7)$ (left) and $Z(17)$ (right) on lattices with several values of $L$.}
\label{Magn_suscet}
\end{center}
\end{figure}  

\begin{figure}
\begin{center}
\includegraphics[scale=0.25]{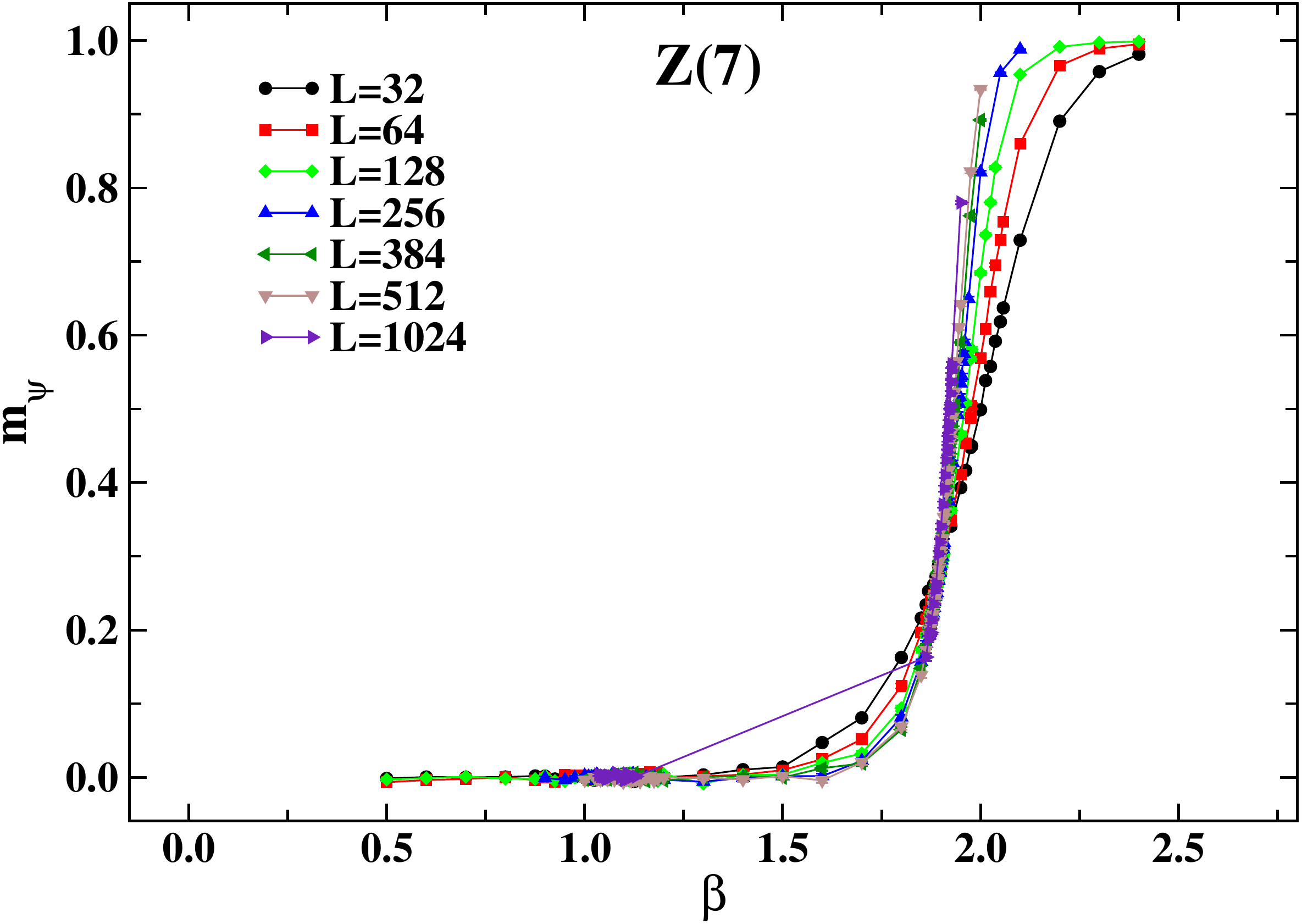}
\hspace{0.5cm}
\includegraphics[scale=0.25]{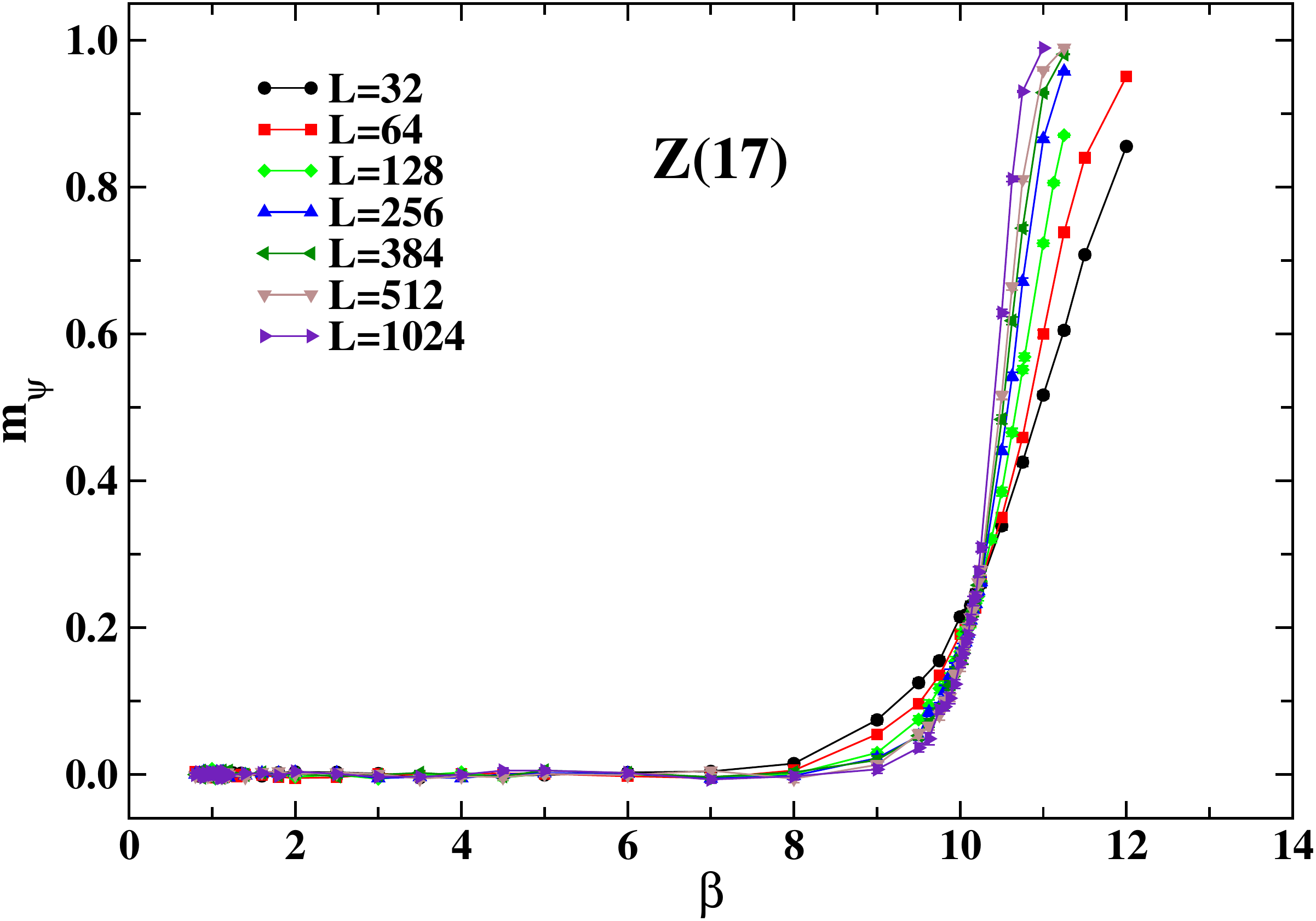}
\end{center}
\vspace{-0.5cm}
\caption{Behavior of $m_{\psi}$ with $\beta$ in $Z(7)$ (left) and $Z(17)$ 
(right) on lattices with several values of $L$.}
\label{Mpsi}
\end{figure}

In Fig.~\ref{Magn_suscet} we show the behavior of the susceptibility 
$\chi_{L}^{(M)}\equiv L^2 (\langle |M_L|^2\rangle - \langle |M_L|\rangle^2)$
of the absolute value of the complex magnetization, which
exhibits, for each volume considered, a clear peak signalling the first phase
transition. The position of the peak in the thermodynamic limit defines 
the first critical coupling, $\beta_{c}^{(1)}$. Fig.~\ref{Mpsi} shows instead 
the behavior of $m_\psi$ versus $\beta$ on various lattice sizes; here the 
second critical coupling $\beta_{c}^{(2)}$ is identified by the crossing 
point (in the thermodynamic limit) of the curves formed by the data on
different lattice sizes. 

To determine the first critical coupling $\beta_{c}^{(1)}$, we could 
extrapolate to infinite volume the pseudo-critical couplings 
given by the position of the peaks of $\chi_{L}^{(M)}$. However, since the 
approach to the thermodynamic limit is rather slow (powers  of $\log L$),
we adopted a different method, based on the use of the 
``\textsl{reduced fourth-order}'' Binder cumulant 
\begin{equation}
U^{(M)}_L=1-\frac{\langle |M_L|^4 \rangle}{3\langle |M_L|^2 \rangle^2} \; ,
\label{binder_U}
\end{equation}
the cumulant $B_4^{(M_R)}$ defined as
\begin{equation}
B_4^{(M_R)}=\frac{\langle |M_R-\langle M_R\rangle|^4\rangle}
{\langle |M_R-\langle M_R\rangle|^2\rangle^2}\;,
\label{binder_MR}
\end{equation}
and the helicity modulus $\Upsilon$.
We estimated $\beta_c^{(1)}$ by looking for (i) the crossing point of the 
curves, obtained on different volumes, giving a Binder cumulant versus 
$\beta$ and (ii) the optimal overlap of the same curves after 
plotting them versus $(\beta - \beta_{c}) (\log L )^{1/\nu}$, with $\nu$ fixed
at 1/2. The method (ii) has been applied also to the helicity modulus 
$\Upsilon$. Our best values for $\beta_{c}^{(1)}$ are
\[
N=7: \;\;\; \beta_{c}^{(1)}=1.1113(13) \;, 
\;\;\;\;\;\;\;\;\;\;\;\;\;\;\;\;\;\;\;\;
N=17: \;\;\; \beta_{c}^{(1)}=1.11375(250)\;.
\]
Then, we performed the finite size scaling (FSS) analysis of the magnetization 
$|M_{L}|$ and the susceptibility $\chi_{L}^{(M)}$ at $\beta_{c}^{(1)}$
using the following laws:
\begin{equation}
|M_{L}|(\beta_{c}^{(1)}) = A L^{-\beta/\nu} \;, \;\;\;\;\;\;\;\;\;\;\;\;\;\;\;
\chi_{L}^{(M)}(\beta_{c}^{(1)}) = B L^{\gamma/\nu}\;,
\label{law_magn}
\end{equation}
where $\gamma/\nu=2-\eta$ and $\eta$ is the {\em magnetic critical index}. 
Results are summarized in Tables~\ref{scaling_magn} 
and~\ref{scaling_suscet_magn}. We observe that the hyperscaling relation 
$\gamma/\nu + 2\beta/\nu = d=2$ is nicely satisfied within statistical errors 
in both models. 

\begin{table}[ht]
\centering
\caption[]{Results of the fit to the data of $|M_{L}|(\beta_{c}^{(1)})$ 
with the scaling law~(\ref{law_magn}-left) on $L^{2}$ lattices with 
$L\geqslant L_{\rm min}$, for $N=7$ and $N=17$.}
\vspace{0.2cm}
\begin{tabular}{|c|c|c|c|}
\hline
\multicolumn{4}{|c|}{$N=7$}\\
\hline
 $L_{\rm min}$ & $A$ & $\beta/\nu$ & $\chi^{2}$/d.o.f. \\
\hline
 32  & 1.00653(48) & 0.12210(08) & 5.5  \\
 64  & 1.00858(70) & 0.12243(12) & 3.7  \\
 128 & 1.01074(94) & 0.12277(15) & 2.0  \\
 256 & 1.0146(16)  & 0.12336(26) & 0.40 \\
 384 & 1.0162(22)  & 0.12359(34) & 0.19 \\
 512 & 1.0177(38)  & 0.12381(56) & 0.16 \\
 640 & 1.0185(57)  & 0.12393(84) & 0.28 \\
\hline
\end{tabular}
\begin{tabular}{|c|c|c|c|}
\hline
\multicolumn{4}{|c|}{$N=17$}\\
\hline
 $L_{\rm min}$ & $A$ & $\beta/\nu$ & $\chi^{2}$/d.o.f. \\
\hline	
  32 & 1.00388(51) & 0.12111(09) & 7.98 \\
  64 & 1.00620(69) & 0.12149(12) & 3.58 \\
 128 & 1.0089(11)  & 0.12191(18) & 1.54 \\
 256 & 1.0107(15)  & 0.12219(24) & 0.74 \\
 384 & 1.0113(24)  & 0.12228(36) & 1.36 \\
\hline
\end{tabular}
\label{scaling_magn}
\end{table}


\begin{table}[ht]
\centering
\caption[]{Results of the fit to the data of $\chi_{L}^{(M)}(\beta_{c}^{(1)})$ 
with the scaling law~(\ref{law_magn}-right) on $L^{2}$ lattices with 
$L\geqslant L_{\rm min}$, for $N=7$ and $N=17$.}
\vspace{0.2cm}
\begin{tabular}{|c|c|c|c|}
\hline
\multicolumn{4}{|c|}{$N=7$}\\
\hline
 $L_{\rm min}$ & $B$ & $\gamma/\nu$ & $\chi^{2}$/d.o.f. \\
\hline
  32 & 0.00558(07) & 1.7402(23) & 2.38   \\
  64 & 0.00540(09) & 1.7457(29) & 1.32   \\
 128 & 0.00522(12) & 1.7508(38) & 0.68   \\
 256 & 0.00518(20) & 1.7520(61) & 0.84   \\
 384 & 0.00546(33) & 1.7443(93) & 0.70   \\	
 512 & 0.00489(52) & 1.760(16)  & 0.28   \\
 640 & 0.00444(76) & 1.775(25)  & 0.0066 \\

\hline
\end{tabular}
\begin{tabular}{|c|c|c|c|}
\hline
\multicolumn{4}{|c|}{$N=17$}\\
\hline
 $L_{\rm min}$ & $B$ & $\gamma/\nu$ & $\chi^{2}$/d.o.f. \\
\hline	
  32 & 0.00559(08) & 1.7372(26) & 2.7  \\
  64 & 0.00532(11) & 1.7453(35) & 0.39 \\
 128 & 0.00521(15) & 1.7484(46) & 0.16 \\
 256 & 0.00514(23) & 1.7507(71) & 0.15 \\
 384 & 0.00522(31) & 1.7483(92) & 0.13 \\
\hline
\end{tabular}

\label{scaling_suscet_magn}
\end{table}

\begin{figure}
\begin{center}
\includegraphics[scale=0.25]{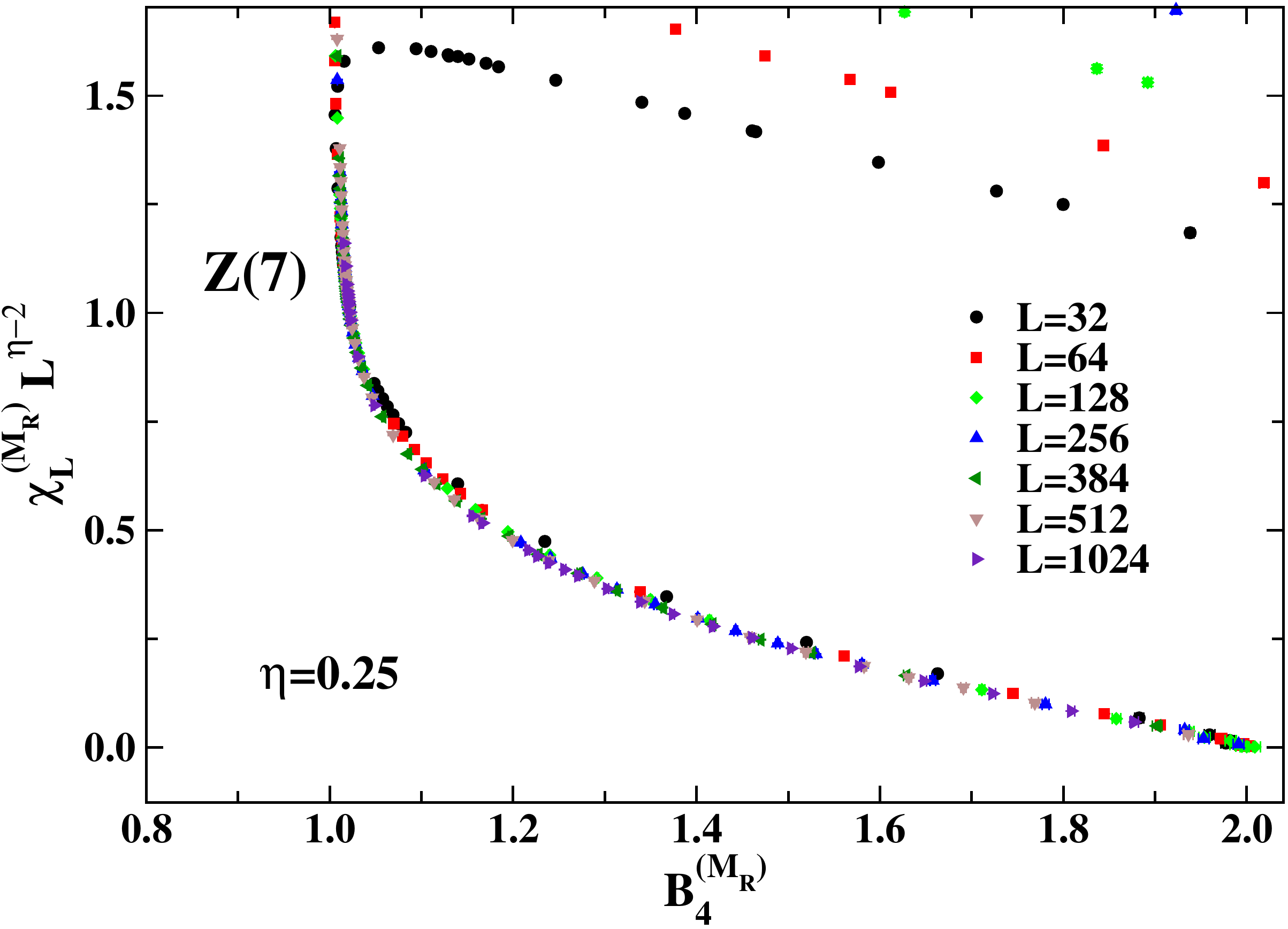}
\hspace{0.5cm}
\includegraphics[scale=0.25]{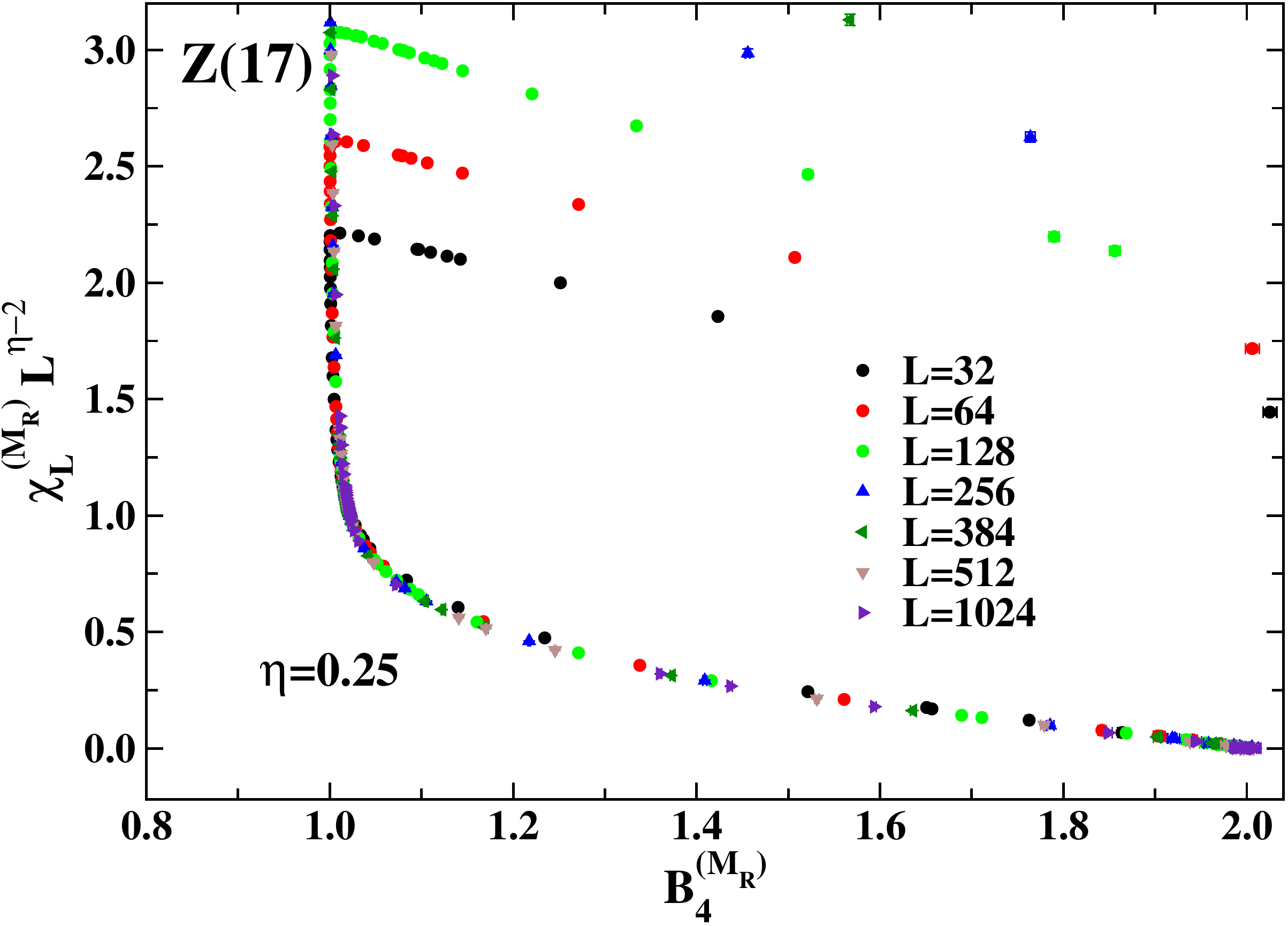}
\end{center}
\vspace{-0.5cm}
\caption{Correlation between $\chi_{L}^{(M_{R})}L^{\eta-2}$ and the Binder 
cumulant $B_{4}^{(M_{R})}$ for $\eta=0.25$ in $Z(7)$ (left) and $Z(17)$ 
(right) on lattices with $L$ ranging from 128 to 1024.}
\label{Susc_rot_b4}
\end{figure}

We can cross-check our determination of the critical exponent $\eta$ by
an independent method, which does not rely on the prior knowledge of
the critical coupling, but is based on the construction of a suitable 
universal quantity~\cite{Loison,z5_phys.rev}. The idea is to plot 
$\chi_{L}^{(M_{R})}L^{\eta-2}$ versus $B_{4}^{(M_{R})}$ and to look for the value of $\eta$ which optimizes the overlap of curves from different volumes.
We found that, both in $Z(7)$ and $Z(17)$, $\eta=1/4$ is this optimal
value, since it gives the best overlap of these curves in the region of 
values corresponding to the first phase transition, {\it i.e.}
the lower branch of the curves of Fig.~\ref{Susc_rot_b4}. This result for 
$\eta$ agrees with the determinations $\eta=2-\gamma/\nu$ from the FSS 
analysis.

As for the second critical coupling $\beta_{c}^{(2)}$, we used the same method
adopted for $\beta_{c}^{(1)}$, but applied now to $B_{4}^{(M_{R})}$ 
and $m_{\psi}$. Our best estimates are
\[
N=7: \;\;\; \beta_{c}^{(2)}=1.8775(75)\;,
\;\;\;\;\;\;\;\;\;\;\;\;\;\;\;
N=17: \;\;\; \beta_{c}^{(2)}=10.13(12)\;.
\]
The standard FSS analysis applied to the susceptibility $\chi_{L}^{(M_{R})}$ 
of the rotated magnetization $M_{R}$ at $\beta_c^{(2)}$ leads to the result
for the critical indices $\gamma/\nu$ given in 
Table~\ref{scaling_magn2}~\footnote{We do not report in this work
the determinations of $\beta/\nu$ by the FSS analysis of the rotated 
magnetization $M_{R}$, since they are affected by large statistical and 
systematic uncertainties.}.

\begin{table}[ht]
\centering
\caption[]{Results of the fit to the data of 
$\chi_{L}^{(M_{R})}(\beta_{c}^{(2)})$ with the scaling 
law~(\ref{law_magn}-right)
on $L^{2}$ lattices with $L\geqslant L_{min}$, for $N=7$ and $N=17$.}
\vspace{0.2cm}
\begin{tabular}{|c|c|c|c|}
\hline
\multicolumn{4}{|c|}{$N=7$}\\
\hline
$L_{min}$ & $A$ & $\gamma/\nu$ & $\chi^2$/d.o.f. \\
\hline	
  32 & 0.8767(37) & 1.92340(71) & 2.02 \\
  64 & 0.8833(47) & 1.92219(87) & 1.41 \\
 128 & 0.8858(57) & 1.9217(11)  & 1.57 \\
 256 & 0.8997(93) & 1.9193(16)  & 1.02 \\
 384 & 0.916(15)  & 1.9166(25)  & 0.68 \\
 512 & 0.921(24)  & 1.9158(39)  & 0.98 \\
 640 & 0.942(34)  & 1.9124(54)  & 1.07 \\
\hline
\end{tabular}
\begin{tabular}{|c|c|c|c|}
\hline
\multicolumn{4}{|c|}{$N=17$}\\
\hline
$L_{min}$ & $B$ & $\gamma/\nu$ & $\chi^2$/d.o.f. \\
\hline
  32 & 0.9319(47)  & 1.98933(89) & 1.65 \\
  64 &  0.9408(68) & 1.9878(12)  & 1.38 \\
 128 &  0.9533(89) & 1.9857(16)  & 0.67 \\
 256 &  0.954(16)  & 1.9856(27)  & 0.83 \\
 384 &  0.950(24)  & 1.9861(40)  & 1.09 \\	
 512 & 0.931(38)   & 1.9892(62)  & 1.44 \\
 640 & 0.911(59)   & 1.9925(98)  & 2.67 \\
\hline
\end{tabular}
\label{scaling_magn2}
\end{table}

\begin{figure}
\begin{center}
\includegraphics[scale=0.25]{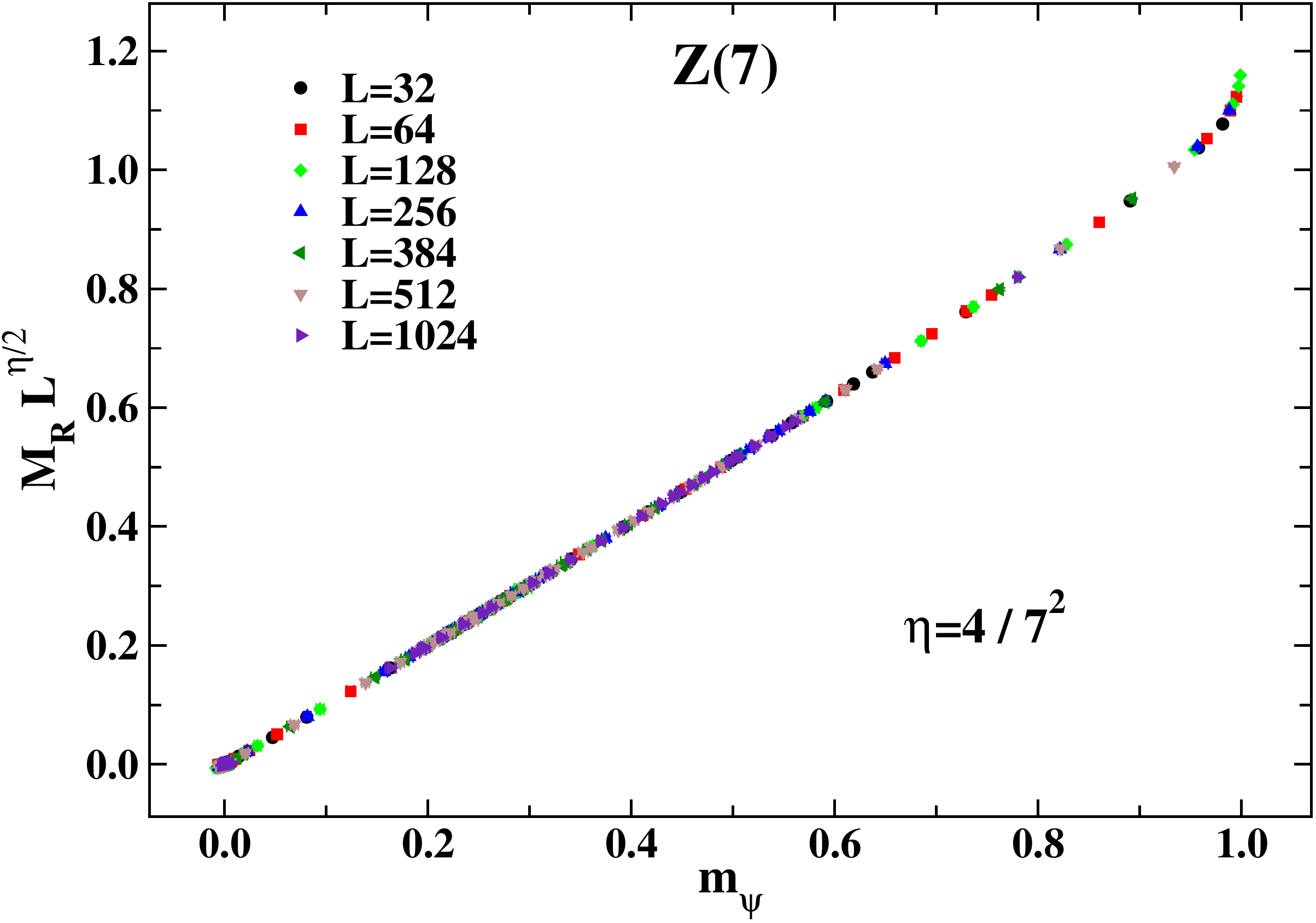}
\hspace{0.5cm}
\includegraphics[scale=0.25]{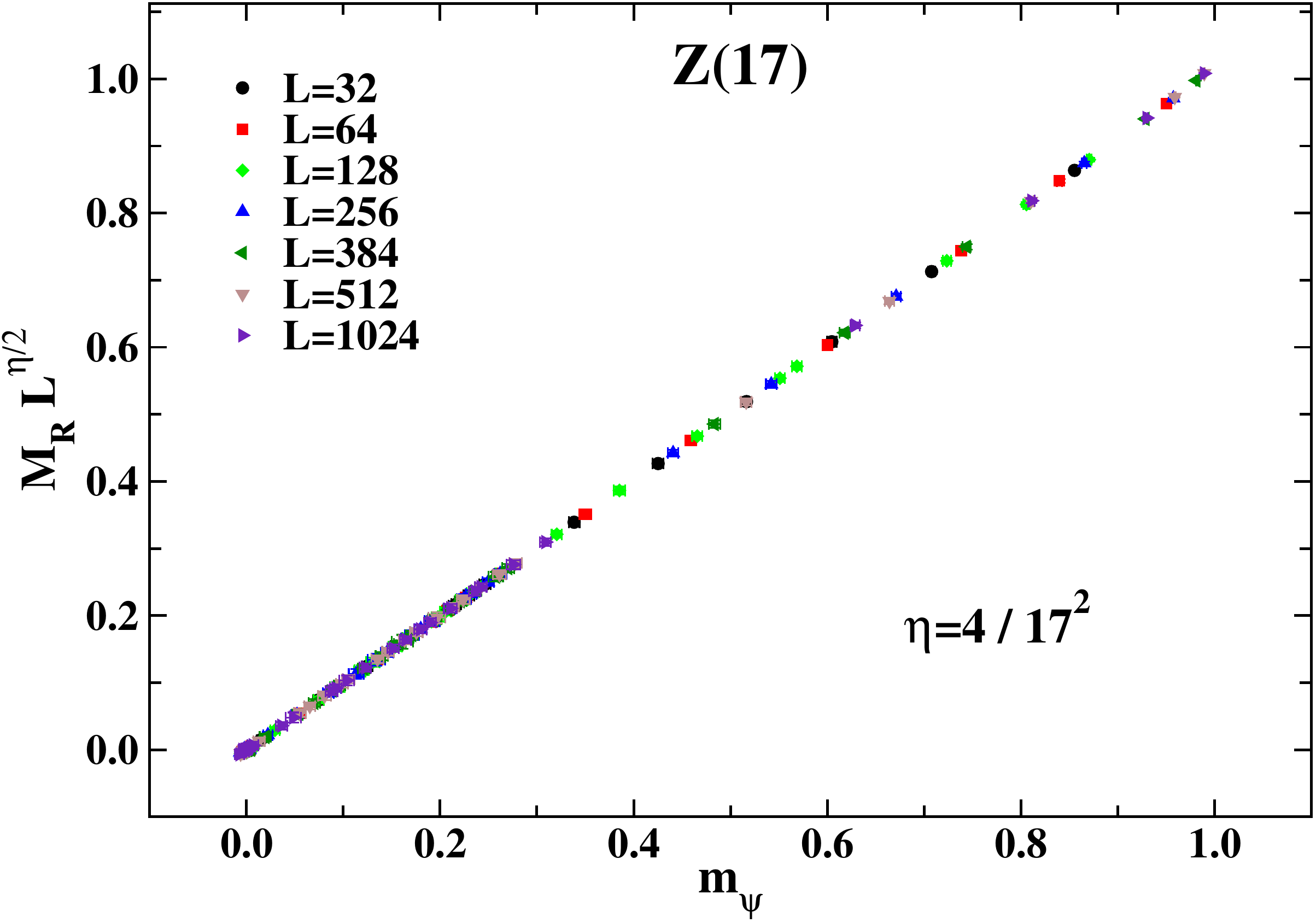}
\end{center}
\vspace{-0.5cm}
\caption{Correlation between $M_{R}L^{\eta/2}$ and $m_{\psi}$ for 
$\eta=4/7^{2}$ in $Z(7)$ (left) and $\eta=4/17^{2}$ in $Z(17)$ (right), on 
lattices with L ranging from 128 to 1024.}
\label{M_mpsi}
\end{figure}

Also in this case the critical index $\eta$ can be determined by an 
independent method, irrespectively of the knowledge of $\beta_c^{(2)}$:
$M_{R}L^{\eta/2}$ is plotted versus $m_{\psi}$ and the value of $\eta$ is
searched for, which optimizes the overlap of data points coming 
from different volumes. The results we found for $\eta$ in $Z(7)$ and $Z(17)$ 
are in perfect agreement with the theoretical prediction $\eta^{(2)}=4/N^{2}$
(see Fig.~\ref{M_mpsi}).

\begin{figure}
\begin{center}
\includegraphics[scale=0.2]{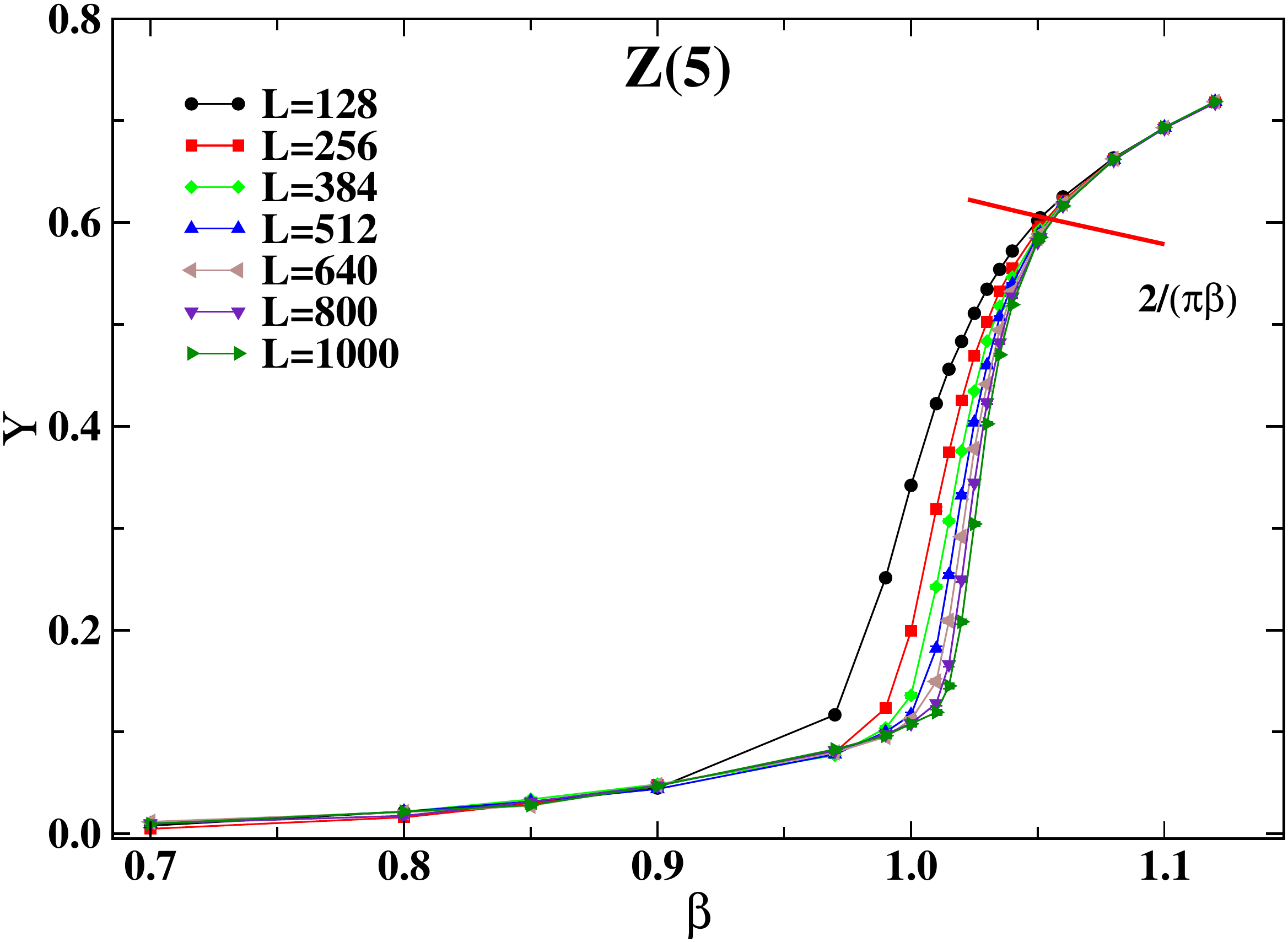}
\includegraphics[scale=0.2]{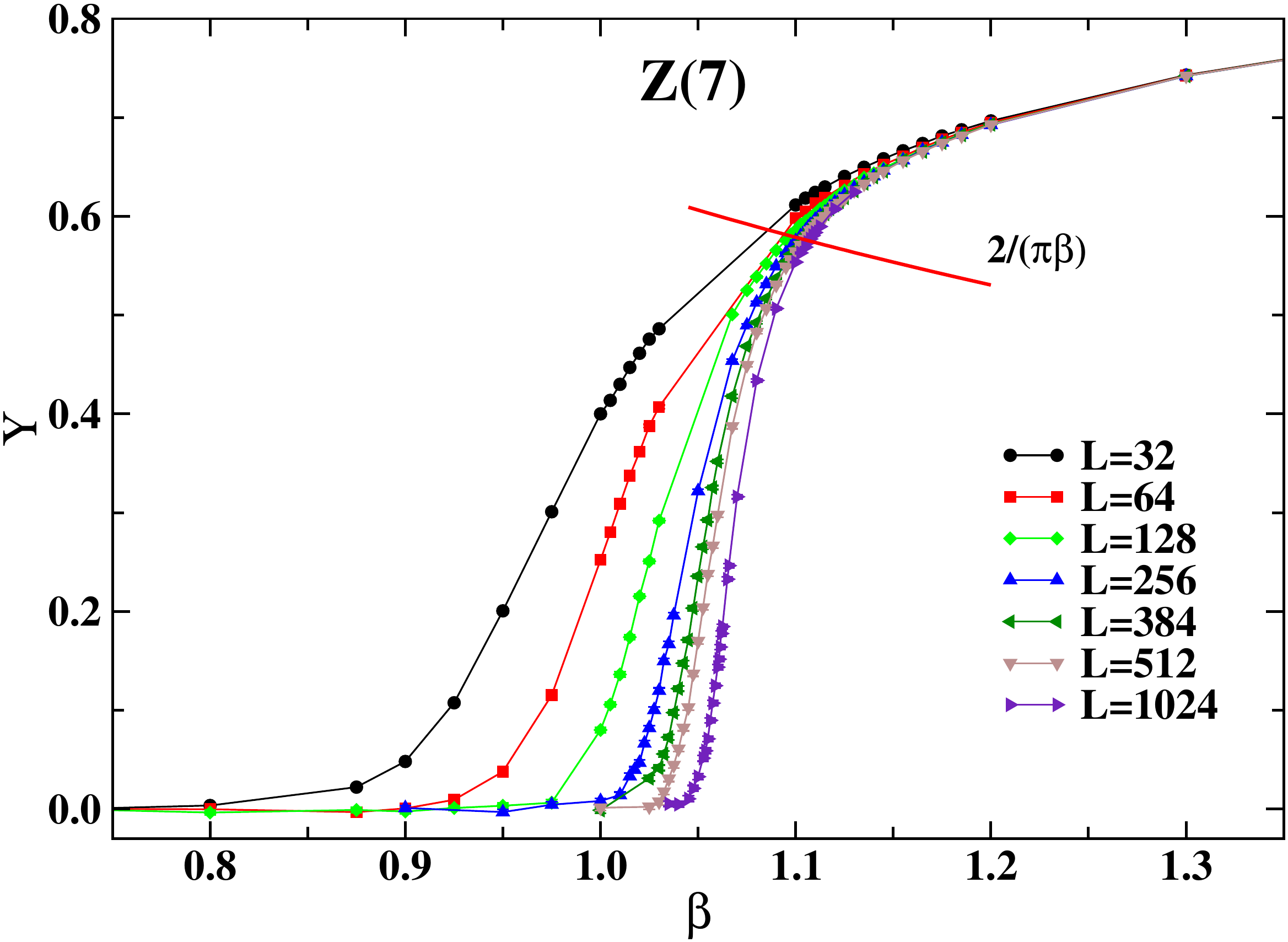}
\includegraphics[scale=0.2]{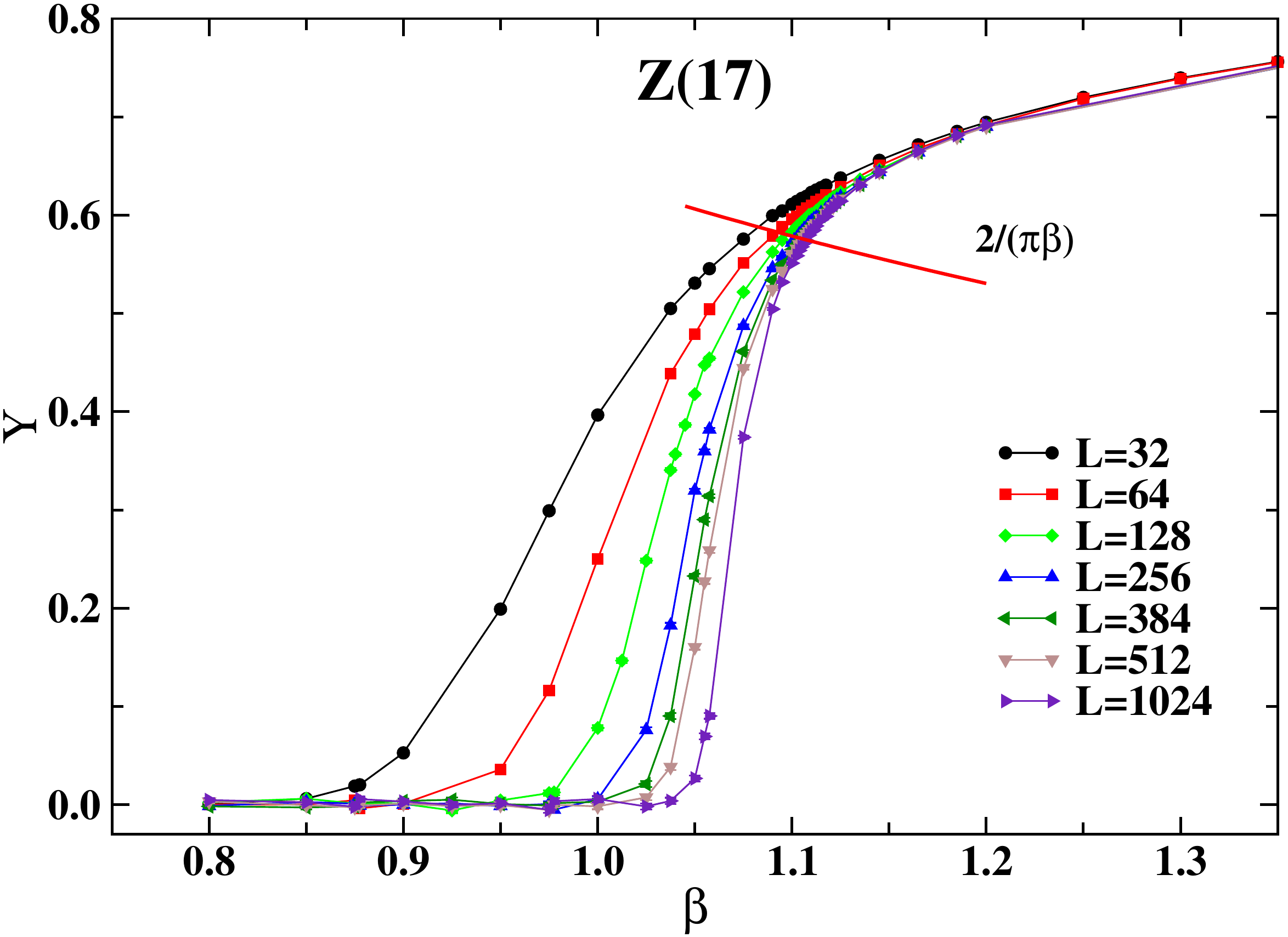}
\end{center}
\vspace{-0.5cm}
\caption{Helicity modulus versus $\beta$ in $Z(5)$, $Z(7)$ and $Z(17)$
on lattices with various sizes.}
\label{Helicity}
\end{figure}

Finally, in Fig.~\ref{Helicity} we present the behavior with $\beta$ of the 
helicity modulus~(\ref{Helicity_formula}). This quantity is constructed in 
such a way that it should exhibit a discontinuous jump (in the thermodynamic 
limit) at the critical temperature separating the disordered phase from the 
massless one, if the transition is of infinite order (BKT). 
Since the Kosterlitz-Thouless RG equations for the $XY$ 
model~\cite{BKT,Ohta,Nelson} 
lead to the prediction that the helicity modulus $\Upsilon$ jumps from the 
value $2/(\pi \beta)$ to zero at the critical temperature, one can check 
if the same occurs for vector Potts models. In Fig.~\ref{Helicity} we plot a 
red line, representing the function $2/(\pi \beta)$; the crossing between this
line and the curves formed by data points of $\Upsilon$ approaches indeed
$\beta_{c}^{(1)}$ when the lattice size increases.

We studied also the specific heat at the two transitions, finding that,
in contrast to the case of first- and second-order phase transitions, 
it does not reflect any nonanalytical critical properties at the critical 
temperatures, thus confirming that only BKT transitions are at work here.  

\section{Summary}

We have determined the two critical couplings of the $2D$
$Z(N=7,17)$ vector models and given estimates of the critical indices $\eta$
at both transitions. Our findings support for all $N\geq 5$ the standard 
scenario of three phases: a disordered phase at high temperatures, a
massless or BKT one at intermediate temperatures and an ordered phase, 
occurring at lower and lower temperatures as $N$ increases. 
This matches perfectly with
the $N\to\infty$ limit, {\it i.e.} the $2D$ $XY$ model, where the ordered 
phase is absent or, equivalently, appears at $\beta\to\infty$. 
Considering the determinations of the critical coupling 
$\beta_c^{(1,2)}(N)$ obtained in this work for $N=7$ and 17, together with 
those obtained for $N=5$ in Refs.~\cite{z5_lat10,z5_phys.rev} and those 
for $N=6$, 8 and 12 of Ref.~\cite{cluster2d}, one can verify that 
$\beta_c^{(1)}(N)$ approaches the $2D$ $XY$ value, $\beta_c^{(1)}=1.1199$,
exponentially in $N$ or even faster, while $\beta_c^{(2)}(N)$ grows to 
infinity with $N^2$. A more detailed analysis of the $N$-behavior of critical 
couplings will be given elsewhere~\cite{noi}.

We have also found that the values of the critical index $\eta$ at the two 
transitions are compatible with the theoretical expectations.

\section{Acknowledgments}

The work of G.C. and M.G. was supported in part by the European Union 
un\-der ITN STRO\-NG\-net (grant PITN-GA-2009-238353).

\end{document}